\documentclass[
    ,final            
  ]
  {aipproc}

\usepackage{bm,graphicx}

\layoutstyle{8x11single}

\begin{document}

\title{Brownian Motion of Stars, Dust, and Invisible Matter}


\author{Edmund Bertschinger}{
  address={Department of Physics, MIT Room 37-602A, 77
Massachusetts Ave., Cambridge, MA 02139} }

\begin{abstract}
Treating the motion of a dust particle suspended in a liquid as a
random walk, Einstein in 1905 derived an equation describing the
diffusion of the particle's probability distribution in
configuration space. Fokker and Planck extended this work to
describe the velocity distribution of the particles.  Their equation
and its solutions have been applied to many problems in nature
starting with the motion of Brownian particles in a liquid.
Chandrasekhar derived the Fokker-Planck equation for stars and
showed that long-range gravitational encounters provide a drag
force, dynamical friction, which is important in the evolution of
star clusters and the formation of galaxies.  In certain
circumstances, Fokker-Planck evolution also describes the evolution
of dark (invisible) matter in the universe.  In the early universe,
the thermal decoupling of weakly interacting massive particles from
the plasma of relativistic leptons and photons is governed by
Fokker-Planck evolution.  The resulting dissipation imprints a
minimum length scale for cosmic density fluctuations.  Still later,
these density fluctuations produce stochastic gravitational forces
on the dark matter as it begins to cluster under gravity.  The
latter example provides an exact derivation of the Fokker-Planck
equation without the usual assumption of Markovian dynamics.
\end{abstract}

\maketitle

\section{Introduction}

While working in 1905 at the Swiss Patent Office in Bern, Albert
Einstein wrote two papers and his PhD thesis on the topic of
Brownian motion \cite{Einstein1,Einstein2,EinsteinPhD}.  These
papers and several later works are translated into English in Ref.\
\cite{EBMbook}. Einstein's work on Brownian motion helped Jean
Perrin confirm the existence of atoms and molecules \cite{Perrin09}
and laid the foundations for stochastic processes in statistical
physics, a topic of great importance a century later.

Brownian motion is the erratic movement of macroscopically small
bodies suspended in a liquid.  The phenomenon was described in 1827
by biologist Robert Brown \cite{Brown1828}.  Brown recognized that
random motion occurred equally for living and nonliving particles.
For six decades the phenomenon attracted little attention, although
several authors suggested that the motion was due not to convective
currents in the liquid but instead might be caused by molecular
collisions.  This hypothesis gained credence with the experiments of
Gouy \cite{Gouy}, which stimulated increased attention to the
subject. Einstein --- and independently Sutherland
\cite{Sutherland05} --- provided the first quantitative theory of
Brownian motion, followed closely by Smoluchowski \cite{Smol06}.

An excellent brief account of the research inspired by Brownian
motion is given in Ref.\ \cite{BM100}. Pais's scientific biography
of Einstein \cite{Pais} is the definitive source for the history and
context of Einstein's work.  The current article focuses on the role
of Brownian motion in the development of kinetic theory and its
application to weakly collisional systems in astrophysics.

\section{Einstein's analysis of Brownian Motion}

Einstein recognized that a system of particles suspended in a liquid
would have osmotic pressure and would undergo diffusion like a
mixture of gases. As he did so often in his major works, he
evaluated the dynamical equations at thermal equilibrium. Consider a
suspension with uniform temperature having a number density gradient
--- for example, a teaspoon of sugar being dissolved in water.
Treated as a fluid, the suspension obeys continuity and momentum
equations.  If the mean free time for collisions with the solvent
atoms is sufficiently short, then the suspension will be in a
quasi-equilibrium state even with nonzero density gradient and fluid
velocity. One-dimensional equilibrium implies
\clearpage
\begin{enumerate}
  \item Force balance: $f-\partial p/\partial x=0$
  \begin{itemize}
    \item Ideal gas law (van 't Hoff): $p=nRT/N_A$ ($p=\hbox{osmotic
      pressure}$, $n=\hbox{number density of suspended particles}$,
      $R=\hbox{gas constant}$, $N_A=\hbox{Avogadro's number}$)
    \item Stokes drag force: $f/n=6\pi\eta av$ ($\eta=\hbox{viscosity}$,
     $a=\hbox{particle radius}$, $v=\hbox{velocity}$)
  \end{itemize}
  \item Number flux balance: $nv-D_x\partial n/\partial x=0$
    ($D_x=\hbox{diffusivity}$)
\end{enumerate}
The pressure force and diffusive flux are both proportional to
$\partial n/\partial x$, which may be eliminated to give the
Einstein-Sutherland relation,
\begin{equation}\label{ESrel}
  D_x=\frac{RT}{6\pi\eta aN_A}\ .
\end{equation}
If $D_x$, $\eta$, and $a$ can be measured, this relation allows the
determination of Avogadro's number.  Modern physicists are so used
to the ideal gas law $p=nkT$ that we forget the atomic origin of
Boltzmann's constant, $k=R/N_A$.

In the limit of negligible advective velocity $v$, the number
density obeys the diffusion equation
\begin{equation}\label{diffusion}
  \frac{\partial n}{\partial t}=D_x\frac{\partial^2n}{\partial
    x^2}\ ,
\end{equation}
whose Green's function solution in an unbounded domain is
\begin{equation}\label{diffsol}
  n(x,t)=\frac{N_0}{\sqrt{4\pi D_xt}}\exp
    \left(-\frac{x^2}{4D_xt}\right)\ .
\end{equation}
We see that diffusion leads to a steadily growing mean square
displacement for each Brownian particle,
\begin{equation}\label{diffwalk}
  \langle x^2\rangle=2D_xt\ .
\end{equation}
This is the same law as a random walk because diffusion is
equivalent to an ensemble of random walks.  A Java applet
illustrating Brownian motion as a random walk is available online at
\url{http://www.aip.org/history/einstein/brownian.htm}.

These results allow one to measure Avogadro's number $N_A$ by
measuring the temperature and viscosity of the solute and the size
of the Brownian particles, and then measuring the diffusivity from
$\langle x^2\rangle(t)$ for Brownian motion \cite{BM06}.  Perrin
achieved the technical breakthrough of uniform-sized Brownian
particles by repeated centrifugation of resin particles over a
period of months. His work on sedimentation and the ``discontinuous
structure of matter'' (i.e. the existence of atoms, in the stilted
English of the Swedish Academy) led to the Nobel Prize in Physics in
1926.

Einstein was unable to measure the size of Brownian particles this
way.  Instead, in his PhD thesis, Einstein cleverly deduced how the
viscosity of a suspension varies with the volume fraction of the
suspended particles, which is proportional to $a^3N_A$.  If the
diffusivity is known, then equation (\ref{ESrel}) leads to an
estimate of $N_A$.  Einstein considered a sugar-water solution,
boldly treating the sugar molecules themselves as Brownian
particles, even though their individual motions were invisible.
Measurements of the diffusivity and viscosity as functions of
concentration led him to an estimate of the sucrose molecule size
and Avogadro's number: $a=6.2$ A, $N_A=3.3\times10^{23}$.  The
calculation contained an arithmetic error in the viscosity
calculation, which was pointed out several years later by a student
of Perrin's. When Einstein corrected the error \cite{Einstein2}, he
obtained $a=4.9$ A, leading to the much more accurate result
$N_A=6.6\times10^{23}$.

\section{Kinetic theory of Brownian Motion}

Brownian motion can be described in two complementary ways: random
walks and diffusion.  The former approach follows individual
particle trajectories while the second follows a distribution
function.  In this section we begin with the Einstein-Smoluchowski
theory of random walks and conclude by exploring diffusion in
velocity space with the Fokker-Planck equation.

The simplest description of Brownian motion is a sequence of
impulses (instantaneous velocity changes) separated by ballistic
motion. After $N$ steps of duration $\Delta t=t/N$ starting from
position ${\bf x}_0$,
\begin{displaymath}
  {\bf x}={\bf x}_0+\sum_{i=1}^N{\bf v}_i\Delta t\ .
\end{displaymath}
Suppose that the velocities have zero mean and are statistically
independent (a Markov process), with
\begin{displaymath}
  \langle{\bf v}_i\rangle=0\ ,\ \
  \langle{\bf v}_i{\bf v}_j\rangle=\sigma_v^2{\bf I}\,\delta_{ij}
\end{displaymath}
where ${\bf I}$ is the unit tensor.  Then by the Central Limit
Theorem, as $N=t/(\Delta t)\to\infty$, ${\bf x}-{\bf x}_0$ becomes
Gaussian with covariance
\begin{equation}\label{posvar}
  \langle({\bf x}-{\bf x}_0)({\bf x}-{\bf x}_0)\rangle=
    N(\sigma_v\Delta t)^2\,{\bf I}=2D_xt\,{\bf I}\ ,\ \
    D_x=\frac{1}{2}\sigma_v^2(\Delta t)\ ,
\end{equation}
recovering the result of equation (\ref{diffwalk}).

Random walks in position are unrealistic because the velocity cannot
change instantaneously.  Assuming a random walk in velocity with
zero-mean, statistically independent accelerations ${\bf a}_i$,
repeating the above derivation gives
\begin{equation}\label{velvar}
  \sigma_v^2=2D_vt\ ,\ \ D_v=\frac{1}{2}\sigma_a^2(\Delta t)\ .
\end{equation}
In thermal equilibrium, $\sigma_v^2=kT/M$ where $M$ is the mass of
the Brownian particle, yielding the unphysical result
$kT=2MD_vt\to\infty$ as $t\to\infty$.  It was invalid to assume zero
mean acceleration. As a Brownian particle's velocity ${\bf v}$
increases, it sees a larger flux of background particles moving
opposite its velocity, and collisions with them transfer a net
momentum proportional to $-{\bf v}$. (Stokes drag is the macroscopic
equivalent.) Assuming a linear drag force and discrete Markovian
dynamics, averaging over the acceleration for a given velocity gives
\begin{equation}\label{accelmom}
  \langle{\bf a}_i\rangle=-\gamma{\bf v}_i\ ,\ \
  \langle{(\bf a}_i-\langle{\bf a}_i\rangle)({\bf a}_j-
    \langle{\bf a}_j\rangle)\rangle
    =\frac{2D_v}{\Delta t}\,{\bf I}\,\delta_{ij}\ .
\end{equation}
Now, ${\bf v}_{i+1}={\bf v}_i+{\bf a}_i (\Delta t)$.  Averaging over
velocities, in equilibrium we must have $\langle{\bf v}_i\rangle
=\langle{\bf v}_{i+1}\rangle=0$ and $\langle{\bf v}_i{\bf v}_i
\rangle=\langle{\bf v}_{i+1}{\bf v}_{i+1}\rangle=(kT/M)\,{\bf I}$.
Working to first order in $\Delta t$, and imposing the Markovian
condition $\langle({\bf a}_i-\langle{\bf a}_i\rangle){\bf v}_i
\rangle=0$, we obtain the important result
\begin{equation}\label{fluctdiss}
  D_v=\gamma\frac{kT}{M}\ .
\end{equation}
This formula is an example of the fluctuation-dissipation theorem:
in equilibrium, diffusivity (describing fluctuations) is
proportional to damping.  At first glance, the Einstein-Sutherland
relation (\ref{ESrel}) seems to violate this because the diffusivity
is inversely proportional to the viscosity.  However, the
Einstein-Sutherland derivation is valid only in the limit of
overdamped motion, for which $D_v=\gamma^2 D_x$ with
$\gamma=6\pi\eta a/M$.

A continuous description of Brownian motion in velocity space was
made possible by the introduction of a stochastic differential
equation by Langevin in 1908 \cite{Langevin}:
\begin{equation}\label{Langeq}
  \frac{d{\bf v}}{dt}=-\gamma{\bf v}+{\bm\Gamma}(t)\ ,
\end{equation}
where ${\bm\Gamma}$ is a stochastic force with mean and covariance
\begin{equation}\label{forcemom}
  \langle\Gamma(t)\rangle=0\ ,\ \
  \langle\Gamma(t)\Gamma(t-t')\rangle=2D_v\delta(t-t')\,{\bf I}\ .
\end{equation}
Langevin multiplied equation (\ref{Langeq}) by ${\bf x}$ and
averaged to derive the Einstein-Sutherland relation, which he showed
is valid only for $\gamma t\gg1$.  However, the Langevin equation
has a much wider applicability to stochastic phenomena in physics
and other disciplines.

The Langevin equation focuses attention on particle trajectories. It
is often more convenient to describe a system statistically using
distribution functions.  A lucid translation between these two
descriptions was provided by Klimontovich \cite{Klim} and Dupree
\cite{Dupree}.  A system of particles has one-particle velocity
distribution function
\begin{equation}\label{dist1}
  f({\bf v},t)=\sum_i\phi({\bf v}-{\bf v}_i(t))\ ,\ \
  \phi({\bf v})=\delta^3({\bf v})\ .
\end{equation}
Here, $i$ labels the particles and the Dirac delta function is a
unit-normalized distribution.  Evolving the system for a time
$\Delta t$ and Taylor expanding this distribution gives
\begin{equation}\label{expdist}
  \frac{\partial f}{\partial t}=\sum_i\left[-\frac{\partial\phi}
    {\partial{\bf v}}\cdot\frac{d{\bf v}_i}{dt}+\frac{\Delta t}{2}
    \frac{d{\bf v}_i}{dt}\cdot\frac{\partial^2\phi}{\partial{\bf v}
    \partial{\bf v}}\cdot\frac{d{\bf v}_i}{dt}\right]+O(\Delta t)^2\ .
\end{equation}
Now taking the ensemble average and using equations (\ref{Langeq})
and (\ref{forcemom}) yields the Fokker-Planck equation \cite{fp}:
\begin{equation}\label{fpeq}
  \frac{\partial f}{\partial t}=\frac{\partial}{\partial{\bf v}}
  \cdot\left(\gamma{\bf v}f+{\bf D}\cdot\frac{\partial f}{\partial
  {\bf v}}\right)\ ,\ \ {\bf D}=\frac{\gamma kT}{M}\,{\bf I}\ .
\end{equation}
This important equation describes diffusion (heating) and drag in
velocity space.  The fluctuation-dissipation theorem --- the linear
relation between diffusivity and drag --- is necessary to ensure the
correct equilibrium solution
\begin{equation}\label{thermeq}
  f({\bf v})=n\left(\frac{M}{2\pi kT}\right)^{3/2}\exp\left(
    -\frac{Mv^2}{2kT}\right)\ .
\end{equation}

When the distribution function depends on position as well as
velocity, the Fokker-Planck equation generalizes to the Kramers
equation \cite{Kramers} (often called Fokker-Planck):
\begin{equation}\label{krameq}
  \frac{Df}{dt}\equiv\frac{\partial f}{\partial t}+{\bf v}\cdot
  \frac{\partial f}{\partial{\bf x}}+{\bf g}\cdot\frac{\partial f}
  {\partial{\bf v}}=-\frac{\partial}{\partial{\bf v}}\cdot
  \left({\bf A}f-\gamma{\bf v}f-{\bf D}\cdot\frac{\partial f}
  {\partial{\bf v}}\right)\ .
\end{equation}
Here ${\bf g}$ is the fluid acceleration, while the quantities ${\bf
A}$ (drift), $\gamma$ (drag), and ${\bf D}$ (diffusivity) are called
transport coefficients.  Often, but not always, they are independent
of ${\bf v}$.  (The distinction between ${\bf g}$ and ${\bf A}$ is
then purely conventional.) Note that $Df/dt$ represents the
transport (advection) of particles in the $({\bf x},{\bf v})$ phase
space along characteristics $d{\bf x}/dt={\bf v}$, $d{\bf v}/dt={\bf
g}$. The quantity ${\bf J}={\bf A}f-\gamma{\bf v}f-{\bf
D}\cdot\partial f /\partial{\bf v}$ is a flux density in velocity
space. The total particle number $\int f\,d^3x\,d^3v$ is conserved.

Equation (\ref{krameq}) and its relatives are called
advection-diffusion equations.  They have widespread applications in
plasma physics, astrophysics, and other disciplines
\cite{Gardiner,Risken}.  Although our analysis began with Brownian
motion, the advection-diffusion equation can describe many systems
in which particle interactions play a dynamical role, including
weakly collisional gases.  The advection-diffusion equation and its
transport coefficients must be derived for each application from
more fundamental dynamics, or justified phenomenologically.

The rest of this article will discuss applications of the
advection-diffusion equation in astrophysics.

\section{Brownian Motion of stars}

Our galaxy contains more than 100 globular clusters, dense balls of
$10^5$ or more stars formed early in the galaxy's history.  Some of
these clusters contain many more X-ray sources than are found among
a comparable number of stars in low-density environments elsewhere
in the galaxy \cite{Pooley}.  The X-rays arise from gas transferred
from a close companion and accreting onto a compact object (neutron
star, white dwarf, or black hole). The implication is that some
globular clusters have many more close binaries (per unit mass of
stars) than the rest of the galaxy. Why?

A plausible answer to this question was suggested more than 30 years
ago \cite{GCXRB}.  Gravitational scattering between a binary and a
third star can transfer energy to the third star, resulting in a
more tightly bound (hence more compact) binary.  If the internal
velocity of the binary is greater than the typical speed of other
stars (i.e., it is a ``hard'' binary), encounters preferentially
remove energy from the binary (causing it to ``harden''). Once a
stellar binary is sufficiently hard, stellar evolution can form a
compact object which accretes from its companion.

When energy is removed from a binary the relative velocities of the
two stars increases.  This is true for any self-gravitating system
close to dynamical equilibrium, as can be seen from the classical
virial theorem.  Let $K$ and $V$ be the total kinetic and potential
energy per unit mass of the cluster, respectively.  The virial
theorem for an inverse square law force states $\langle2K+V
\rangle\approx0$ where the angle brackets denote a time average over
a timescale longer than the characteristic time for stars to cross
the cluster. Using the virial theorem we can evaluate the
temperature-dependence of the specific heat. The total energy per
unit mass is $E=K+V$ and the ``temperature'' is proportional to the
kinetic energy per unit mass.\footnote{The velocity distribution for
a system of bodies with only gravitational forces is generally
non-Maxwellian.}  The specific heat is then
\begin{equation}\label{specheat}
  \frac{dE}{dT}\propto\frac{d}{dK}(K+V)\approx-\frac{dK}{dK}=-1
\end{equation}
where the virial theorem has been used.  Self-gravitating systems
have a negative specific heat and are therefore thermodynamically
unstable \cite{LBW}.  This is true for any central force with
two-body potential proportional to $r^n$ with $-2<n<0$.  It is also
true for black holes in general relativity (for which the
temperature is the Hawking temperature.)

The thermodynamic instability operates on the timescale for particle
interactions to thermalize the cluster. Suppose that the core of a
cluster shrinks slightly.  As a consequence of the virial theorem
the core heats up.  Scattering by gravitational interactions
conducts heat outward, causing the core to lose energy and thereby
contract further.  The outer part of the cluster expands with the
addition of energy. Heat conduction in a stellar dynamical system is
a diffusive process governed approximately by the Fokker-Planck
equation \cite{Chandra}.

Qualitatively, the kinetic theory for a stellar dynamical system
describes fluctuations caused by two-body scattering superposed on
mean-field dynamics.  The rate for two-body scattering can be
estimated using the Rutherford (Coulomb) cross section for
scattering of bodies with relative speed $v$:
\begin{equation}\label{scattrate}
  \gamma\sim n\sigma v\ln\Lambda\ ,\ \ \sigma\approx\left(
    \frac{GM}{v^2}\right)^2\ ,
\end{equation}
where we have assumed equal mass stars; the ``Coulomb logarithm''
$\ln\Lambda$ accounts for long-range interactions, which dominate
the total cross section for momentum transfer.  The drag coefficient
$\gamma\propto v^{-3}$ for gravitational interactions; for a nearly
Maxwellian distribution with velocity dispersion $\sigma_v$ the
velocity-space diffusivity $D\approx\gamma \sigma_v^2$. For the
densest globular clusters, $\gamma^{-1}\sim10^9$ yr is much less
than the age of the clusters, implying that two-body scattering has
had ample time to drive the collapse of the core and, plausibly, the
formation of compact binaries which then become X-ray sources.

\begin{figure}[t]
    \includegraphics[scale=0.8]{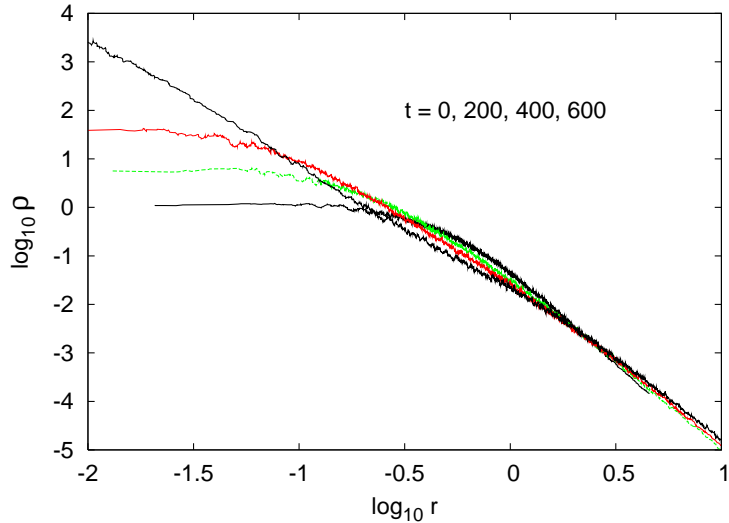}
  \caption{Spherically averaged density profile of a $N=1000$ star
  cluster (the average of 5 clusters) evolved starting from a Plummer
  sphere with core radius $r=3\pi/16$.  (The total mass and energy of
  the cluster are 1 and $-1/4$, respectively; the crossing time is
  $t_{\rm cr}=2\sqrt{2}$ and the relaxation time is $t_{\rm rel}=0.1N
  t_{\rm cr}/\ln N=41$.)  Two-body scattering transfers heat out from
  the core of the cluster causing it to shrink and become increasingly
  dense until hard binaries form.}
  \label{fig:farr}
\end{figure}

Modern computing power makes it almost possible (with approximate
treatments of stellar evolution and tidal interactions) to study
globular cluster dynamics directly using full numerical simulation
\cite{HeggieHut}. A more modest example illustrating the
gravothermal instability is shown in Figure \ref{fig:farr} using
exact pairwise forces and a fourth-order symplectic integrator
\cite{farrbert}.

\section{Dark Matter and Brownian Motion in the Early Universe}

Most of the mass in the universe is in some form other than atomic
(``baryonic'') matter.  This matter is invisible --- it neither
scatters nor absorbs detectable amounts of electromagnetic radiation
--- and has been dubbed ``dark matter.'' Dark matter has been
detected only through its gravitational effect on atomic matter and
light.  Its existence has been suspected for 70 years and known with
confidence for about 25 years.  However, the exact nature of dark
matter is still unknown.

The most natural hypothesis is that dark matter consists of
particles or fields not yet detected in the laboratory. Astrophysics
and cosmology provide strong constraints on the nature of this
substance.  The particles (or field excitations) must have
sufficiently small thermal speed to provide the seeds for galaxy
formation.  This rules out standard model neutrinos.  It cannot be
atomic matter in any form without having overproduced helium during
big bang nucleosynthesis.  While proposals have been made that dark
matter is an illusion arising from modified gravity \cite{MOND}, the
only known theoretically and experimentally consistent
implementation of modified gravity without dark matter
\cite{Bekenstein} is far less economical and less well tested than
the dark matter hypothesis.

Dark matter is natural in extensions of the standard model
\cite{particleDM}, with two leading candidates: axions and WIMPs
(weakly interacting massive particles).  Axions are hypothetical
ultra-low energy excitations of the $\theta$-vacuum of QCD that
arise naturally as a solution of the strong CP problem (i.e., why
the strong interactions conserve CP, or why the neutron has a tiny
electric dipole moment). They are spin-0 particles of mass $10^{-6}$
to $10^{-2}$ eV and despite their light mass have negligible thermal
speeds because they form a Bose-Einstein condensate.

WIMPs are popular dark matter candidates because supersymmetry and
other extensions of the standard model of particle physics call for
particles with approximately the correct mass and cross section to
produce the observed abundance of dark matter \cite{LeeWeinberg}.
The favored dark matter candidate is the lightest neutralino
$\chi^0$, the supersymmetric partner of a linear combination of the
photon, $Z^0$, and Higgs bosons.  The neutralino is
spin-$\frac{1}{2}$, has mass $m_\chi$ in the range 20 to 500 GeV,
and is its own antiparticle.

At present, these dark matter models are speculative.  Proof will
come only from direct detection of dark matter particles in the
laboratory.  Nonetheless, the models are sufficiently compelling to
merit detailed examination of their astrophysical consequences.  The
remainder of this section focuses on WIMPS as they are better
studied than axions.

WIMP dark matter may be effectively collisionless today, but in the
early universe WIMPs scattered rapidly with fermions in the
relativistic plasma.  Inelastic processes such as $\chi^0\chi^0
\leftrightarrow L\bar L$, where $L$ is a lepton, maintained the
abundance of WIMPs in chemical equilibrium until the reaction rates
fell below the Hubble expansion rate (chemical decoupling or
freezeout). This happened after the WIMPs became nonrelativistic and
their abundance decreased exponentially through annihilation.  At a
temperature $\sim m_\chi/25$, annihilation ceased to be effective
and the abundance of WIMPs froze out to
$n(\chi^0)/n(L)\sim10^{-10}$.

After annihilation ceased, WIMPS continued to undergo elastic
scattering $\chi^0 L \leftrightarrow\chi^0 L$ with abundant leptons
until the temperature had fallen to a few MeV (not coincidentally,
slightly before neutrino decoupling at 2 MeV).  During this period
the WIMPs underwent Brownian motion as they were bombarded by the
light relativistic leptons, with two important consequences. First,
the WIMPs were maintained in kinetic (thermal) equilibrium with the
relativistic plasma until the elastic scattering rate fell below the
Hubble expansion rate (kinetic decoupling). Second, friction between
the WIMPs and leptons damped density perturbations generated in the
early universe.  WIMP dark matter should therefore have a thermal
cutoff in its perturbation spectrum at length scales corresponding
roughly to the Hubble distance at kinetic decoupling.

The Brownian motion of WIMPs can be studied starting from the
relativistic Boltzmann equation.  Expanding the Boltzmann collision
integral in powers of the momentum transfer divided by the WIMP
mass, for nonrelativistic WIMPs one obtains the relativistic
Fokker-Planck equation
\begin{eqnarray}\label{relFP}
  \frac{Df}{d\tau}&\equiv&g^{\mu\nu}\left(p_\mu\frac{\partial f}
    {\partial x^\nu}+\Gamma^\lambda_{\ \,i\mu}p_\lambda p_\nu
    \frac{\partial f}{\partial p_i}\right)\nonumber\\
  =\left(\frac{df}{d\tau}\right)_c&=&\gamma\frac{\partial}{\partial
    {\bf p}}\cdot\left[({\bf p}-m_\chi{\bf v}_L)f+m_\chi kT_L
    \frac{\partial f}{\partial{\bf p}}\right]\ ,
\end{eqnarray}
where $f({\bf x},{\bf p},t)$ is the distribution function in a local
Lorentz frame, ${\bf v}_L$ and $T_L$ are the mean fluid velocity and
temperature of the lepton fluid, and $\gamma$ is the collision rate
coefficient determined from particle physics.

\begin{figure}[t]
    \includegraphics[scale=0.8]{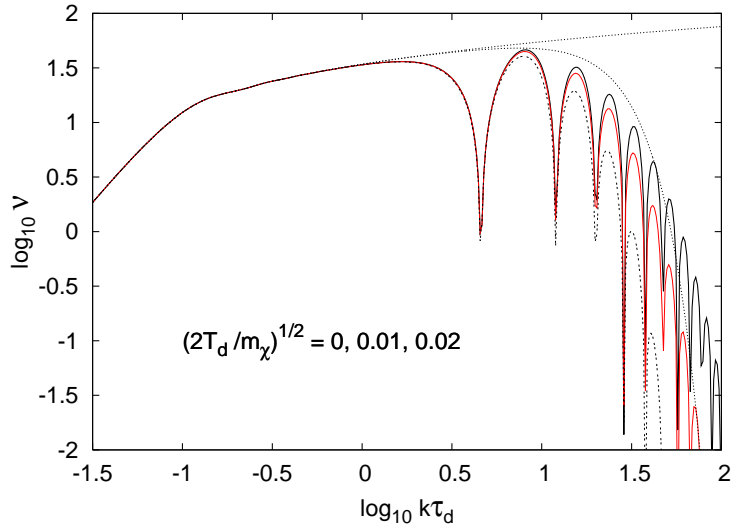}
  \caption{Cold dark matter density transfer function versus wavenumber
    (in units of the inverse decoupling time) after kinetic decoupling
    in the early universe; $\nu$ is the factor by which primordial
    fluctuations grow due to gravitational instability. Here $T_d$ is
    the temperature at which kinetic decoupling takes place; realistic
    models have $0.01<(2T_d/m_\chi)^{1/2}<0.02$ (larger values of this
    parameter give more damping). The upper, monotonic curve assumes,
    unrealistically, that the dark matter is always collisionless and
    was never coupled to the radiation.  The other non-oscillating curve
    shows a crude model of  kinetic decoupling described by a Gaussian
    cutoff.}
    \label{fig:chi}
\end{figure}

Equation (\ref{relFP}) and the Einstein and fluid equations for the
photon-lepton plasma have been solved numerically  including the
acoustic oscillations of the plasma before and during kinetic
decoupling, the frictional damping occurring during kinetic
decoupling, and the free-streaming damping occurring afterwards and
throughout the radiation-dominated era \cite{b06}.  For a
$m_\chi=100$ GeV WIMP with bino-type interactions, kinetic
decoupling occurs at a temperature $T_d=16$ MeV.  Figure
\ref{fig:chi} shows the net effect of WIMP-lepton scattering on the
linear transfer function of cold dark matter density perturbations.
The damped oscillations are analogous to the acoustic peaks of the
cosmic microwave background anisotropy and the baryon acoustic
oscillations imprinted on the galaxy distribution.  However, because
WIMPs decouple at a redshift $z\sim10^{10}$ instead of $z=1300$, the
length scale of these fluctuations is about 1 pc instead of 100 Mpc.
This length scale corresponds to a mass $10^{-4}\,(T_d/$10
MeV)$^{-3}$ M$_\odot$.  It is possible, though at the moment
controversial, that substructures as small as this might persist in
our galaxy today and be detectable through the products of rare dark
matter annihilations \cite{diemand}.

\section{Brownian Motion in Galaxy Formation}

After kinetic decoupling in the early universe, dark matter dynamics
is governed entirely by gravity.  As shown above, stars undergo
Brownian motion in a self-gravitating cluster.  However, dark matter
particles are much lighter than stars and their two-body interaction
time is far longer than the age of the universe.  Nonetheless, dark
matter particles can scatter from quasiparticles or substructure in
galaxy halos much as electrons scatter from substructure in the
nucleon during deep inelastic scattering.  In effect, galaxy halos
are filled with gravitational partons.

The concept of dark matter substructure is based on the fact that
nonlinear gravitational instability accumulates mass into dense
clumps conventionally called dark matter halos (or minihalos or
microhalos for the very small ones) \cite{ps74}.  In the
hierarchical clustering paradigm for structure formation,
small-scale structures form first and then aggregate into larger
objects where they persist for some time.  Gravitational N-body
simulations have shown that the spherically-averaged radial density
profiles of the evolved halos take a nearly universal form
\cite{nfw} without providing an explanation for this result. One
possibility is that scattering by substructure leads to relaxation.

The BBGKY hierarchy provides the exact statistical description for
the evolution of a classical gas.  The first BBGKY equation is
similar to the Boltzmann equation except that it is exact, not
phenomenological, and it is not closed:
\begin{equation}\label{bbgky1}
  \frac{\partial f}{\partial t}+{\bf v}\cdot\frac{\partial f}
    {\partial{\bf x}}+{\bf g}\cdot\frac{\partial f}{\partial
    {\bf v}}=-\frac{\partial}{\partial{\bf v}}\cdot{\bf F}_v\ ,\ \
    \frac{\partial}{\partial{\bf x}}\cdot{\bf F}_v=4\pi G
    \rho_{2c}({\bf x},{\bf v},t)\ .
\end{equation}
Here $\rho_{2c}$ is an integral over the phase-space two-point
correlation function.  The second BBGKY equation gives the evolution
of the two-point correlation in terms of the three-point
correlation, etc. For weakly correlated gases, these higher-order
correlations may be neglected, and the two-point correlation term
may be approximated by a Boltzmann collision integral.  A different
approach is needed for a gravitational plasma, where Boltzmann's
Stosszahlansatz does not apply.

In the early stages of gravitational clustering, $\rho_{2c}$ can be
evaluated using second-order cosmological perturbation theory and
the first BBGKY equation reduces to a Fokker-Planck equation
\cite{mabert04}, with surprising results.  First, the diffusivity
eigenvalues can be negative, which appears to be a consequence of
gravitational instability.  Second, in the quasilinear regime there
is no drag --- $\gamma=0$ in equation (\ref{krameq}) --- but there
is a nonzero radial drift ${\bf A}=A(r)\vec e_r$.  Dynamical
friction arises only in higher orders of perturbation theory (or in
the fully nonlinear regime).  The radial drift is induced by
substructure.  Finally, the timescale for relaxation obtained from
the drift and diffusivity is the Hubble time, i.e. the collapse time
of the initial perturbation.  This initial relaxation is
surprisingly fast.

The only approximation made in deriving this Fokker-Planck equation
for dark matter evolution was second-order perturbation theory. The
starting point was the exact BBGKY equation, not the
phenomenological master equation, and there was no assumption of
Markov dynamics.  Indeed, the time evolution of a given realization
of the random process is completely smooth.  Why, then, is the
system described by a Fokker-Planck equation?  The answer is that
the equation of motion for dark matter particles is a modified
Langevin equation:
\begin{equation}\label{cosmolang}
  \frac{d{\bf v}}{dt}=-2H(t){\bf v}+b(t){\bf g}\ ,
\end{equation}
where $H(t)$ is the Hubble expansion rate, $b(t)$ is proportional to
the growth factor of density perturbations, and ${\bf g}({\bf x})$
is a Gaussian random field. For one realization of this process
(i.e., one universe) this field is definite and the acceleration of
every particle is smooth. The kinetic equation governs the evolution
of the average dark matter halo, hence involves averaging over an
ensemble of Gaussian random fields. This averaging leads to a
Fokker-Planck equation.

The derivation suggests that spatial fluctuations of the density
field can cause relaxation, but is not conclusive because the most
important dynamical effects occur only in the fully nonlinear
regime. Dynamical friction of substructure (both as ``field'' and
``test'' particles) and tidal stripping of substructure must be
incorporated into the description.  A major stumbling block is the
statistical characterization of the substructure and its effects.
Most investigations of substructure neglect velocity information,
giving an inadequate description of phase space correlations. It
remains to be seen whether a satisfactory kinetic theory can be
devised for the fully nonlinear regime of dark matter gravitational
clustering.

\section{Conclusions}

Brownian motion continues to serve as a paradigm for stochastic
processes in physics and other disciplines.  Einstein's great
insight was that the dynamics of Brownian motion can be understood
by applying thermodynamic equilibrium to the interaction between a
fluid of Brownian particles and the fluid in which those particles
are suspended.

The universality of thermodynamics arises because most statistical
systems near equilibrium relax at rates calculable from thermal
equilibrium. Self-gravitating systems like globular star clusters
and galaxies, with their negative specific heats and lack of
thermodynamic equilibrium, are an important exception. While N-body
simulations are the main tool for studying the evolution of
self-gravitating systems, analytical insight can be obtained from
kinetic theory, especially from the diffusive evolution driven by
fluctuations.

A century after Einstein's analysis of Brownian motion, the kinetic
theory of self-gravitating systems remains a largely unsolved
problem, presenting a great opportunity for a future Einstein --- or
a Fokker, Kramers, or Dupree.

\begin{theacknowledgments}
I thank Jean-Michel Alimi for organizing and hosting a wonderful
conference and Will Farr for providing Figure \ref{fig:farr}. This
work was supported by the Kavli Foundation and by National Science
Foundation grant AST-0407050.
\end{theacknowledgments}

\bibliographystyle{aipproc}   

\end{document}